\documentstyle[12pt,aaspp4,flushrt]{article}

\slugcomment{Submitted to The Astrophysical Journal}

\catcode`\@=11 
\def\@versim#1#2{\vcenter{\offinterlineskip
        \ialign{$\m@th#1\hfil##\hfil$\crcr#2\crcr\sim\crcr } }}
\newcommand{\beq}{\begin{equation}}
\newcommand{\eeq}{\end{equation}}
\def\lsim{\mathrel{\mathpalette\@versim<}}
\def\gsim{\mathrel{\mathpalette\@versim>}}

\begin{document}
\title{On angular momentum transport in convection-dominated
accretion flows}

\author{Igor V. Igumenshchev}

\affil{Laboratory for Laser Energetics, University of Rochester, 250
East River Road, Rochester, NY 14623; iigu@lle.rochester.edu}

\medskip

\begin{abstract}

Convection-dominated accretion flow (CDAF) is a promising model
to explain underluminous accreting black holes in X-ray binaries
and galactic nuclei.
I discuss effects of angular momentum transport in viscous
hydrodynamical and MHD CDAFs.
In hydrodynamical CDAFs, convection transports
angular momentum inward, and this together with outward convection
transport of thermal energy determine the radial structure of the flow.
In MHD CDAFs, convection can transport
angular momentum either inward or outward, depending on properties
of turbulence in rotating magnetized plasma, which are not fully
understood yet. Direction of convection angular momentum transport
can affect the law of rotation of MHD CDAFs.

\noindent {\it Subject Headings:} accretion, accretion disks ---
black hole physics --- convection --- hydrodynamics --- MHD ---
turbulence

\end{abstract}

\section{Introduction}

A phenomenon of underluminous accreting solar mass black holes in
X-ray binaries and supermassive black holes in galactic nuclei has
stimulated recent investigations of radiatively inefficient
accretion flows (see Narayan, Mahadevan \& Quataert 1998 and Narayan
2002 for reviews). Contrary to a standard Shakura-Sunyaev accretion disk
model, which successfully explains relatively soft and luminous
X-ray sources,
models of radiatively inefficient flows are used to explain
the significant deficit of radiation observed in some hard X-rays sources.
A particular example of such underluminous sources
is the Galactic center, Sagittarius~$A^*$, which hosts a $2\times 10^6 M_\odot$
black hole. The Galactic center has a luminosity that is well below
estimates based on the models of Shakura-Sunyaev accretion 
disk or spherical Bondi accretion flow (Melia \& Falcke 2001).
Other examples of underluminous supermassive black holes have
been provided by recent Chandra observations of ellipticals NGC~1399,
NGC~4472, NGC~4636 and NGC~6166 
(Loewenstein et al. 2001; di Matteo et al. 2001).

Numerical simulations of radiatively inefficient accretion flows,
made in the framework of viscous hydrodynamical (HD) approach, revealed
that low viscosity flows are convectively unstable and
convection strongly influences structure of such flows (Igumenshchev,
Chen, \& Abramowicz 1996; Igumenshchev \& Abramowicz 1999, 2000;
Stone, Pringle, \& Begelman 1999; Igumenshchev, Abramowicz, \& Narayan
2000; McKinney \& Gammie 2002).
An analytical model, which reproduces basic features of the HD
simulations, was called convection-dominated accretion flow (CDAF,
see Narayan, Igumenshchev, \& Abramowicz 2000; Quataert \& Gruzinov 2000).
CDAFs consist of a hot plasma at about virial temperatures and have 
a flatten time-averaged radial density profile, $\rho\propto R^{-1/2}$,
where $\rho$ is the density and $R$ is the radius.
In CDAFs, a significant fraction of binding
energy, most of which is released in the innermost region of accretion
flows, is transported outward by convection motions. 
HD CDAFs rotate with about Keplerian angular velocity (somewhat
below the Keplerian value), and
angular momentum is simultaneously transported
inward by convection
and outward by anomalous viscosity, so the net angular momentum
flux is nearly zero (Narayan et al. 2000).

HD models of rotating accretion flows
use {\it ad hoc} assumption on the nature of angular momentum transport.
Specifically, one must assume a non-zero viscosity $\nu$ to provide
outward angular momentum transport in the flows; without such a transport
inward spiraling of matter into the black hole is not possible.
An important disadvantage of the HD approach is that
functional dependences of $\nu$ on other parameters of the problem
can not be self-consistently found. Moreover, it is not even clear
that a HD viscous diffusion model of angular momentum transport
could correctly describe
real flows. It is believed that the problem can be
self-consistently solved in the frame-work of MHD approach.
A small-scale MHD turbulence (with characteristic length-scale of magnetic
field $\ell < R$)
initiated by the magneto-rotational instability (Balbus \&
Hawley 1991) and/or large-scale magnetic loops ($\ell \ga R$) 
could be responsible for angular momentum transport in rotating
flows. 

The problem of MHD accretion involves non-linear physical effects,
requiring numerical simulations.
Global MHD numerical simulations
of radiatively inefficient accretion flows have been started recently
by several research groups, thanks to development of high performance
computers and advances in numerical technics.
Two groups have used similar designs of numerical experiments, 
in which calculations have been started with an equilibrium torus
orbiting a black hole.
Despite of this similarity, the designs are importantly different in
initial configuration of magnetic field.
One group ( Machida, Matsumoto, \& Mineshige 2001) 
assumed that the torus has only
toroidal component of magnetic field at the beginning of simulations. 
Evolution of the torus after several orbital periods
was resulted in development of small-scale
MHD turbulence and convection motions; 
inner parts of the torus formed a turbulent accretion flow. 
Machida et al. (2001) noted that the accretion flow has 
radial structure similar to CDAFs found in the low viscosity HD simulations.
Another group
(Stone \& Pringle 2001; Hawley 2000; 
Hawley, Balbus, \& Stone 2001; Hawley \& Balbus
2002) assumed poloidal magnetic field inside the tori as
an initial condition of their numerical models.
Resulting accretion flows were significantly different from 
the results of Machida et al. (2001).
These flows formed a powerful bipolar magnetized outflows,
which sandwich a relatively narrow region of equatorial inflow.
The bipolar outflows efficiently
remove angular momentum from the inflowing material, providing accretion. 
Convection was not observed in these simulations.
The numerical results of these two groups have demonstrated
that topology of magnetic field
in rotating accretion flows is an important factor determined
the flow structure, and mechanisms and efficiency of angular momentum 
transport.

Study of spherical (non-rotating) MHD accretion flows 
provided further insight on the role of convection in radiatively
inefficient accretion flows. It was shown with help of 3D
MHD numerical simulations (Igumenshchev \& Narayan 2002)
that magnetized Bondi flows 
experience a transition to convection-dominated regime of accretion.
This regime (convection-dominated Bondi flow or CDBF) 
has no HD counterparts among non-rotating
accretion flows, but it is similar in many aspects 
to the rotating HD CDAFs.
Unless different physical mechanisms are responsible for
development of convection in HD CDAF and CDBF, 
simulations showed that convection itself
governs the flow structure independently on details of those mechanisms.
This explains a formal similarity of analytic self-similar solutions
described HD CDAFs and CDBFs (see Narayan et al. 2001; Igumenshchev \&
Narayan 2002). 

In this Letter I will discuss properties of CDAF with
magnetic field, and will
concentrate on the problem of angular momentum transport.
This study is motivated by recent works by Balbus
and collaborators (Hawley et al. 2001; Balbus \& Hawley 2002),
who argued that there is a fundamental difference in dynamics
of MHD and viscous HD flows. Because of this difference and
consequent difference in mechanisms of angular momentum transport,
they concluded that CDAFs can not be formed with magnetized plasma
(for discussions, see Narayan et al. 2002).
I will show, based on analytic arguments and
results of previous numerical simulations,
that CDAFs can exist independently of the changes of details of 
angular momentum transport, provided that the changes are 
due to magnetic field.


\section{Self-similar solutions}

Analytical theory of CDAF is based on a self-similar solution
of a simplified set of equations described radiatively inefficient
accretion flows. I briefly reproduce here main results of the theory
following to Narayan et al. (2000) and Igumenshchev \& Narayan (2001).
Consider stationary height-integrated equations described
an accretion flow onto the black hole of mass $M$.
In absence of mass outflows, the continuity equation reads
$$ \dot{M}=-4\pi RH\rho v_R, \eqno (2.1) $$
where $\dot{M}$ is the accretion rate,
$H=(c_s/v_K)R$ is the scale height,
$c_s=\sqrt{P/\rho}$ is the sound speed,
$P$ is the pressure,
$v_R$ is the accretion velocity, $v_R<0$,
$v_K=\sqrt{GM/R}$ is the Keplerian velocity and 
$G$ is the gravitational constant.

In the radial momentum equation, 
I ignore the convection term $v_R dv_R/dR$,
assuming that $v_R^2\ll v_K^2$.
In addition, I ignore the turbulent
pressure and consider the total pressure $P$ consisted of the gas 
pressure $P_g=R_g\rho T$ and magnetic pressure $P_m=P_g/\beta$,
$P=P_g+P_m=(\beta+1)P_g/\beta$, where 
$T$ is the temperature,
$R_g$ is the gas constant, and
$\beta$ is a parameter.
Then the equation takes the form
$$  \Omega^2_KR-\Omega^2R = -{1\over\rho}{dP\over dR}, \eqno (2.2) $$
where $\Omega$ is the angular velocity and $\Omega_K=v_K/R$ is the
Keplerian angular velocity.
Equation (2.2) describes a balance between gravitational, centrifugal
and pressure gradient forces.

The angular momentum equation can be written in the form of balance
of advection and diffusion transport terms,
$$  {1\over R}{d\over dR}\left(R\Sigma v_R {\cal L}\right)=
{1\over R}{d\over dR}\left(\Sigma\nu_{out}R^3{d\Omega\over dR}\right) +
{1\over R}{d\over dR}\left(\Sigma\nu_{in}R{d{\cal L}\over dR}\right),
 \eqno (2.3) $$
where $\Sigma=2\rho H$ is the surface density and ${\cal L}=\Omega R^2$ is the
specific angular momentum.
The term on the left hand side of
(2.3) describes inward advection of angular momentum, whereas the two terms
on the right hand side correspond to two different mechanisms of
diffusion angular momentum transport. The first term
on the right hand side of (2.3), involved the
viscosity coefficient $\nu_{out}$, describes the standard viscous
transport, when the corresponded angular momentum flux is proportional to
the gradient of angular velocity. In stationary accretion flows,
$\Omega$ is always decreasing function of radius, so 
the term describes outward transport of angular momentum.
In the second term on the right hand side of (2.3),
the angular momentum flux is proportional to the gradient of
angular momentum with the viscosity coefficient $\nu_{in}$. This term 
describes a tendency of axisymmetric turbulence to uniform
distribution of angular momentum in accretion flows 
(e.g. Ryu \& Goodman 1992).
In flows with nearly Keplerian rotation, $\Omega\propto R^{-3/2}$,
the latter term represents inward transport of angular momentum.

The energy equation is
$$  \rho v_R T {ds\over dR}\equiv \rho v_R \left({1\over\gamma-1}
{dc^2_s\over dR}-{c^2_s\over\rho}{d\rho\over dR}\right)=
-{1\over R^2}{d\over dR}\left(R^2F\right)+Q,
 \eqno (2.4) $$
where $s$ is the specific entropy, $\gamma$ is
the adiabatic index,
$F$ is the outward energy flux due to
convection, and $Q$ is the energy dissipation rate per unit volume.
Equation (2.4) states that the locally released internal energy (the term $Q$)
can be advected inward (the term involved $ds/dR$) and/or transported
outward by convection (the term involved $F$).

In the considered theory,
simple parametric expressions are used to represent the viscosity
coefficients $\nu_{out}$ and $\nu_{in}$,
$$ \nu_{out}= \alpha_{out}c_s H,
 \eqno (2.5) $$
$$ \nu_{in}= \alpha_{in}c_s H,
 \eqno (2.6) $$
the convective flux,
$$ F=-\alpha_c c_s R \rho T {ds\over dR},
 \eqno (2.7) $$
and the dissipation rate per unit volume,
$$ Q=-\alpha_{d}{v_R\over R}\rho{GM\over R},
 \eqno (2.8) $$
where $\alpha_{out}$, $\alpha_{in}$, $\alpha_c$ and $\alpha_{d}$ are
positive dimensionless parameters.

Equations (2.1)-(2.4) with (2.5)-(2.8)
have two types of self-similar solutions, 
which correspond to two values of the self-similar index $a=3/2$ and
$1/2$. The solutions can be written as follows,
$$ \rho(R)=\rho_0 R^{-a},
 \eqno (2.9) $$
$$ v_R(R)=-v_0 v_K \left({R_G\over R}\right)^{3/2-a} \propto R^{a-2},
 \eqno (2.10) $$
$$ c_s(R)=c_0 v_K\propto R^{-1/2},
 \eqno (2.11) $$
$$ \Omega=\Omega_0 \Omega_K \propto R^{-3/2},
 \eqno (2.12) $$
where $R_G=2GM/c^2$ is the gravitational radius and
$c$ in the speed of light.
The dimensionless coefficients $v_0$, $c_0$, and $\Omega_0$
in (2.10)-(2.12)
can be expressed algebraically through the parameters
$\alpha_{out}$, $\alpha_{in}$, $\alpha_c$, $\alpha_{d}$, $\beta$ and
$\gamma$, substituting (2.9)-(2.12) into (2.2)-(2.4).
Note, that not all values of the parameters lead to a consistent
solution.
The coefficient $\rho_0$ in (2.9) is proportional to the accretion 
rate $\dot{M}$
and scales out of the problem.
The solution with $a=3/2$ corresponds to advection-dominated accretion
flow (e.g. Narayan \& Yi 1994; Abramowicz et al. 1995), whereas
the solution with $a=1/2$ represents CDAF (Narayan et al. 2000;
Quataert \& Gruzinov 2000). Convection-dominated solution with $a=1/2$
and $\Omega_0=0$ describes CDBF (Igumenshchev \& Narayan 2002).

\section{Angular momentum transport}

Consider the CDAF self-similar solution.
In this case one can neglect the advection terms (the terms involved $v_R$)
in (2.3)-(2.4). Taking into account (2.5) and (2.6),
equation (2.3) can be re-written in the form,
$$ {d\ln\Omega\over d\ln R}=
-{2\alpha_{in}\over\alpha_{out}+\alpha_{in}}.
 \eqno (3.1) $$

HD CDAFs only exists in the form of flows with nearly Keplerian rotation, 
$\Omega\propto\Omega_K$. From this fact and from equation (3.1),
one obtains the following relation between parameters $\alpha_{in}$ and
$\alpha_{out}$,
$$ \alpha_{out} = {1\over 3}\alpha_{in}.  \eqno (3.2) $$
The relation (3.2) significantly limits properties of
convection turbulence in rotating CDAFs: an intensity of axisymmetric
convection motions, which transport angular momentum inward,
must be supported at a some specific level, given by (3.2),
to compensate viscous outward spread of angular momentum.
Could this be truth for real flows?
HD simulations of radiatively inefficient accretion flows
revealed that only CDAF-type solutions exist in the low viscosity
limit.
These solutions have nearly Keplerian rotation,
$\Omega\propto\Omega_K$, 
and convection in these flows is almost axisymmetric due to
ability of the Keplerian shear motion to uniform 
convective perturbations in $\phi$-direction (azimuthal direction).
It was found with help of 2D and 3D HD simulations
(Igumenshchev \& Abramowicz 2000; Igumenshchev et al. 2000)
that the time-averaged $R\phi$-component of Reynolds
stress tensor has negative sign in CDAFs, which confirms that
axisymmetric convection transports angular momentum inward.
Moreover, strengths of inward and outward angular momentum
fluxes were estimated to be close to each other in the numerical models.
Thus, results of numerical simulations of HD CDAFs are in a good qualitative 
and quantitative agreement with results of the self-similar theory
(for more detailed comparison of the theory and simulations, see
Igumenshchev \& Abramowicz 2000).

Magnetic fields can affect
properties of CDAFs found in HD simulations.
As it was discussed in introduction,
there is an important difference between HD and MHD models of 
radiatively inefficient accretion
flow that MHD models have a non-rotating counterpart of CDAFs,
whereas HD models do not have it.
So, in MHD flows there is no longer the requirement
of nearly Keplerian rotation, and
in principle, the accretion flow can consist of
rotating and non-rotating parts with transition regions between them.
Solutions described such transition regions can not be
self-similar and they are out of scope of this Letter.
However, one can better understand properties of MHD CDAFs 
considering flows of extreme rotation rates.

Consider slowly rotating ($\Omega\ll\Omega_K$) MHD CDAFs,
in which rotation has negligible dynamical effect. 
Radial structure of such flows
can be described by the CDBF self-similar solution.
The law of rotation in the flows can be found with help of
(3.1), if one knows the parameters $\alpha_{out}$ and $\alpha_{in}$.
Relative importance of the parameters can be estimated as follows.
In slowly rotating flows, a weak rotation shear is not
able to uniform non-axisymmetric perturbations in $\phi$-direction, 
as it is in the case of fast rotating HD CDAFs,
and convection motions are unlikely to be axisymmetric.
Thus, both non-axisymmetric convection 
and small scale magnetic turbulence
transport  angular momentum outward much more efficiently, 
than inward, and one gets  $\alpha_{out}\gg \alpha_{in}$.
In this limit, using (3.1), one finds that the flow must rotate
as a rigid-body, $\Omega(R)=\Omega_B$, where
$\Omega_B$ is the angular velocity at the outer boundary
(according to our assumption, $\Omega_B\ll\Omega_K$).
Relative dynamical importance of rotation in such slowly rotating CDAFs
is decreased with decreasing of the radius.

Next, consider MHD CDAFs which rotates with about Keplerian
angular velocity at the outer boundary, $\Omega_B\approx\Omega_K(R_B)$.
Two possible structures of the flow can be proposed
in this case, depending on the mechanism of angular momentum transport.
The first structure is identical to the structure of
HD CDAF, in which the flow rotates with nearly Keplerian angular velocity
everywhere down to the inner boundary, $\Omega\propto\Omega_K$.
This requires that
convection turbulence in MHD CDAFs has the same properties
as in the case of HD CDAFs: axisymmetric convection transports angular
momentum inward and almost compensates 
outward angular momentum flux
due to small scale MHD turbulence. 
In such rotating MHD CDAFs,
$\alpha_{in}$ and $\alpha_{out}$ satisfy to (3.2).
The second structure is formed if
the outward angular momentum flux exceeds the inward one.
In this case one has (see eq.~[3.1])
$$ \alpha_{out}> {1\over 3} \alpha_{in} \qquad {\rm and} \qquad
{d\ln\Omega\over d\ln R} > 
{d\ln\Omega_K\over d\ln R}. \eqno (3.3) $$
The excessive flux 
results in formation of a region inside $R_B$,
in which the flow
slows down rotation with respect to Keplerian one,
as it follows from (3.3).
In the limit $R\ll R_B$, 
when the rotation is reduced significantly,
one obtains
the slowly rotating MHD CDAF considered above.
Thus, the second possible structure
of MHD CDAFs can consist of two parts: the outer
transition part, where the flow rotation is slowed down 
from the boundary level at $R_B$
till significantly sub-Keplerian level at $R\ll R_B$, 
and the inner part, where the flow
rotates slowly, almost as a rigid body.

\section{Conclusion}

The convection-dominated accretion flow (CDAF) model consistently
represents radiatively inefficient accretion flows into black holes
in the frame-work of viscous hydrodynamical (HD) approach. 
In HD CDAFs,
rotation is nearly Keplerian, supported by simultaneous
and balanced transport of angular momentum inward by convection and
outward by viscosity.
It is not yet clear how the MHD approach could change the CDAF model.
Results of direct numerical simulations of magnetized radiatively inefficient
accretion flows indicate that solutions strongly depend 
on assumed topology of magnetic field in the flow
(Machida et al. 2001; Hawley \& Balbus 2002). 
At present, however,
absence of detailed MHD simulations makes it difficult to describe
actual structure of magnetized radiatively inefficient accretion flows.

In this Letter I have shown that the existence and properties
of CDAF do not
crucially depend on direction of convection angular momentum transport
in magnetized plasma, contrary to what was thought previously
based on results of HD simulations.
MHD CDAF can exist even if convection presumably transports 
angular momentum outward, rather than inward. 
In this case of dominant outward transport
the flow at small radii becomes similar to non-rotating
convection-dominated Bondi flow (CDBF).
I have considered two possible structures of MHD CDAF, which are
determined by the relative
strengths of inward and outward angular momentum transport.
One possibility is that the radial structures and rotation profiles
of MHD and HD CDAFs are similar,
supposing that
convection in magnetized accretion flows transports angular momentum
inward with the same efficiency as it does in viscous HD flows.
If, however, the inward transport is less efficient
in MHD flows, than the outward one, then 
the radial structure of CDAFs can be different.
In the latter case the flow
consists of two parts. In the outer part, the angular velocity is
reduced from the outer boundary value (supposed to be close to Keplerian
one) till significantly sub-Keplerian value.
In the inner part, the flow rotates very slowly,
so its radial structure is represented by the 
self-similar CDBF solution.
A choice between these two possible structures of MHD CDAFs can be
made on the base of high-resolution 3D numerical simulations.


\acknowledgements

Author thanks Ramesh Narayan and Marek Abramowicz for helpful
discussions and support.
The study was supported by 
the U.S. Department of Energy (DOE) Office of Inertial 
Confinement Fusion under Cooperative Agreement No. DE-FC03-92SF19460,
the University of Rochester, the New York State Energy Research
and Development Authority, and RFBR grant 00-02-16135.
The support of the DOE does not constitute an endorsement by the DOE
of the views expressed in this article.


\end{document}